\documentclass[twoside,twocolumn,9pt]{article}
\usepackage{extsizes}
\usepackage[super,sort&compress,comma]{natbib} 
\usepackage[version=3]{mhchem}
\usepackage[left=1.5cm, right=1.5cm, top=1.785cm, bottom=2.0cm]{geometry}
\usepackage{balance}
\usepackage{times,mathptmx}
\usepackage{sectsty}
\usepackage{graphicx} 
\usepackage{lastpage}
\usepackage[format=plain,justification=justified,singlelinecheck=false,font={stretch=1.125,small,sf},labelfont=bf,labelsep=space]{caption}
\usepackage{float}
\usepackage{fancyhdr}
\usepackage{fnpos}
\usepackage[english]{babel}
\usepackage{array}
\usepackage{droidsans}
\usepackage{charter}
\usepackage[T1]{fontenc}
\usepackage[usenames,dvipsnames]{xcolor}
\usepackage{setspace}
\usepackage[compact]{titlesec}
\usepackage{hyperref}
\usepackage{dblfloatfix}

\usepackage{amsfonts}
\usepackage{amsmath}
\usepackage{amssymb}
\usepackage{bm}
\usepackage{braket}
\usepackage{siunitx}
\usepackage{mathtools}
\usepackage{tikz}
\usepackage{subcaption}

\DeclareSIUnit\Molar{\textsc{m}}

\tikzstyle{place}=[circle,draw,fill,thick,inner sep = 0pt,minimum size = 1.5mm]
\tikzstyle{theline}=[line width = 1pt]

\DeclareSymbolFont{Letters}{OML}{cmm}{m}{it}
\DeclareMathSymbol{\psi}{\mathalpha}{Letters}{32}

\definecolor{cream}{RGB}{222,217,201}
\begin{document}
\pagestyle{fancy}
\thispagestyle{plain}
\fancypagestyle{plain}{

%%%HEADER%%%
%\fancyhead[C]{\includegraphics[width=18.5cm]{head_foot/header_bar}}
%\fancyhead[L]{\hspace{0cm}\vspace{1.5cm}\includegraphics[height=30pt]{head_foot/journal_name}}
%\fancyhead[R]{\hspace{0cm}\vspace{1.7cm}\includegraphics[height=55pt]{head_foot/RSC_LOGO_CMYK}}
\renewcommand{\headrulewidth}{0pt}
}

%%%PAGE SETUP 
\makeFNbottom
\makeatletter
\renewcommand\LARGE{\@setfontsize\LARGE{15pt}{17}}
\renewcommand\Large{\@setfontsize\Large{12pt}{14}}
\renewcommand\large{\@setfontsize\large{10pt}{12}}
\renewcommand\footnotesize{\@setfontsize\size{7pt}{10}}
\makeatother

\renewcommand{\thefootnote}{\fnsymbol{footnote}}
\renewcommand\footnoterule{\vspace*{1pt}% 
\color{cream}\hrule width 3.5in height 0.4pt \color{black}\vspace*{5pt}} 
\setcounter{secnumdepth}{5}

\makeatletter 
\renewcommand\@biblabel[1]{#1}            
\renewcommand\@makefntext[1]% 
{\noindent\makebox[0pt][r]{\@thefnmark\,}#1}
\makeatother 
\renewcommand{\figurename}{\small{Fig.}~}
\sectionfont{\sffamily\Large}
\subsectionfont{\normalsize}
\subsubsectionfont{\bf}
\setstretch{1.125} 
\setlength{\skip\footins}{0.8cm}
\setlength{\footnotesep}{0.25cm}
\setlength{\jot}{10pt}
\titlespacing*{\section}{0pt}{4pt}{4pt}
\titlespacing*{\subsection}{0pt}{15pt}{1pt}

%%%FOOTER%%%
\fancyfoot{}
\fancyfoot[RO]{\footnotesize{\sffamily{1--\pageref{LastPage} ~\textbar  \hspace{2pt}\thepage}}}
\fancyfoot[LE]{\footnotesize{\sffamily{\thepage~\textbar\hspace{2pt} 1--\pageref{LastPage}}}}
\fancyhead{}
\renewcommand{\headrulewidth}{0pt} 
\renewcommand{\footrulewidth}{0pt}
\setlength{\arrayrulewidth}{1pt}
\setlength{\columnsep}{6.5mm}
\setlength\bibsep{1pt}

%%%FIGURE SETUP 
\makeatletter 
\newlength{\figrulesep} 
\setlength{\figrulesep}{0.5\textfloatsep} 

\newcommand{\topfigrule}{\vspace*{-1pt}% 
\noindent{\color{cream}\rule[-\figrulesep]{\columnwidth}{1.5pt}} }

\newcommand{\botfigrule}{\vspace*{-2pt}% 
\noindent{\color{cream}\rule[\figrulesep]{\columnwidth}{1.5pt}} }

\newcommand{\dblfigrule}{\vspace*{-1pt}% 
\noindent{\color{cream}\rule[-\figrulesep]{\textwidth}{1.5pt}} }

\makeatother

%%%%%%%%%%%%%%%%%%%%%%%%%%%%%%%%%%%%%%%%%%%%%%%%%%%%%%
%%%TITLE SECTION %%%%%%
\twocolumn[
  \begin{@twocolumnfalse}
\vspace{3cm}
\sffamily
\begin{tabular}{m{4.5cm} p{13.5cm} }
%%%%%%%%% TITLE %%%%%%%%%
\includegraphics{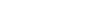} & \noindent\LARGE{\textbf{Polymorphism of stable collagen fibrils}} \\
\vspace{0.3cm} & \vspace{0.3cm} \\
 & \noindent\large{Samuel Cameron,\textit{$^{a}$} Laurent Kreplak,\textit{$^{a}$} and Andrew D. Rutenberg\textit{$^{a}$}} \\
 %%%%%%%%%%% ABSTRACT %%%%%%%%%%
 %
\includegraphics{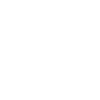} & \noindent\normalsize{Collagen fibrils are versatile self-assembled structures that provide mechanical integrity within mammalian tissues. The radius of collagen fibrils vary widely depending on experimental conditions \textit{in vitro} or anatomical location \textit{in vivo}. Here we explore the variety of thermodynamically stable fibril configurations that are available. We use a liquid crystal model of radial collagen fibril structure with a double-twist director field. Using a numerical relaxation method we show that two dimensionless parameters, the ratio of saddle-splay to twist elastic constants $ k_{24}/K_{22}$ and the ratio of surface tension to chiral strength $\tilde{\gamma} \equiv \gamma/(K_{22}q)$, largely specify both the scaled fibril radius and the associated surface twist of equilibrium fibrils. We find that collagen fibrils are the stable phase with respect to the cholesteric phase only when the reduced surface tension is small, $\tilde{\gamma} \lesssim 0.2$.  Within this stable regime, collagen fibrils can access a wide range of radii and associated surface twists. Remarkably, we find a maximal equilibrium surface twist of $\SI{0.33}{\radian}$ ($\SI{19}{\degree}$).  Our results are compatible with corneal collagen fibrils, and we show how the large surface twist is needed to explain the narrow distribution of corneal fibril radii. Conversely, we show how small surface twist is required for the   thermodynamic stability of tendon fibrils in the face of considerable polydispersity of radius.} \\
\end{tabular}
 \end{@twocolumnfalse} \vspace{0.6cm}
  ]  %%%%% END OF TITLE SECTION %%%%% 
  
%%%FONT SETUP - 
\renewcommand*\rmdefault{bch}\normalfont\upshape
\rmfamily
\section*{}
\vspace{-1cm}

\footnotetext{\textit{$^{a}$~Dept. of Physics and Atmospheric Science, Dalhousie University, Halifax, Nova Scotia, Canada B3H 4R2.}}

%%%%%%%%%%%% INTRODUCTION %%%%%%%%%%%
\section{Introduction}
Tropocollagen is the most abundant protein in the human body, integral to the structure of fibrous tissues such as skin, tendon, and cornea. There are at least 28 different tropocollagen molecules found in vertebrates \cite{Kadler:2007dl}, with types I, II, III, V, XI, XXIV, XXVII capable of forming the rope-like mesostructures that are collagen fibrils \cite{Mouw:2014bs}. The assembly of tropocollagen molecules into collagen fibrils depends on the local environment. A suitable environment {\em in vivo} is within the extra-cellular space \cite{Birk:1986tu} after procollagen, a precursor to tropocollagen, is secreted from cells and cleaved by enzymes \cite{Stephens:2012js}. {\em In vitro}, ionicity, pH, and temperature of the solvent \cite{Wood:1960ko,Harris:2013dv}, as well as concentration \cite{Mosser:2006da}, have been shown to affect whether fibrillogenesis occurs. 

Since fibrils are approximately cylindrical with radius $R$, it is convenient to separately consider their axial structure along the fibril's cylindrical axis and their radial structure within a circular cross section. The axial  D-banding has been well studied and remains close to $\SI{67}{\nano\meter}$ for both type I \cite{Fang:2013gb} and type II \cite{Antipova:2010} collagen. Conversely, the observed radial ultrastructure of collagen fibrils depends on both the tropocollagen type and the anatomic location of the fibril {\em in vivo} \cite{Raspanti:interjbiomacromol1989, Ottani:micron2002, Gutsmann:biophys2003, Lillie:jultrares1977}, and on solution conditions {\em in vitro} \cite{Wood:1960ko}. Factors such as temperature and ionicity of solution \cite{Wood:1960ko}, or fibril age {\em in vivo} \cite{Raspanti:interjbiomacromol1989, Parry:1978bb}, affect the observed fibril radii.

In this work, we focus on radial structure. We are concerned with what constrains the fibril radius, $R$, but also with the orientation of collagen molecules both on the fibril surface and within the fibril.  

Observing the orientation of molecules on the {\em interior} of a circular cross section of fibril is difficult experimentally, requiring diffraction studies \cite{Folkhard:1987gx} or electron tomography \cite{Holmes:2001bt}. However, careful high resolution imaging can reliably characterize the molecular orientation at the {\em surface} of fibrils. Early work using transmission electron microscopy to image freeze-fractured fibrils found that molecules at the fibril surface were tilted with respect to the fibril axis, with the degree of tilt depending on where the fibrils were found anatomically \cite{Ruggeri:1979hw,Reale:1981}. Further work demonstrated that tendon fibrils, while exhibiting a large range of $R$ values (\SI{15}{\nano\meter}-\SI{200}{\nano\meter}), have limited molecular surface tilt $\simeq\SI{5}{\degree}$ \cite{Baselt:1993fk,Hulmes:biophys1995}, while corneal fibrils, with a narrower range of $R$ from $\SI{15}{\nano\meter}-\SI{20}{\nano\meter}$, exhibit much larger surface tilt, $\simeq \SI{18}{\degree}$ \cite{YAMAMOTO:2000di, Hirsch:2001gp, Holmes:2001bt}.  Different hypotheses of radial molecular orientation have been proposed to fit these experimental results \cite{Raspanti:interjbiomacromol1989, Holmes:2001bt, Hulmes:biophys1995} -- but without consideration of thermodynamic stability. 

Recently, an equilibrium liquid crystal model of radial collagen fibril structure was developed to predict molecular configurations of tropocollagen within individual fibrils \cite{Brown:softmatter2014}. Consistent with the surface tilt observations mentioned above, a double-twist geometry of tropocollagen molecules was imposed within the fibril. With this double-twist geometry, the twist angle of molecules with respect to the fibril axis at a given radial distance, $\psi(r)$, fully describes the molecular orientation of the tropocollagen molecules. The corresponding elastic free energy functional \cite{Ferrarini:liqcryst2010}, valid for arbitrary smoothly varying $\psi(r)$, is parameterized by the costs of twist distortion, $K_{22}$, bend distortion, $K_{33}$, saddle-splay distortion, $k_{24}$, surface tension, $\gamma$, and the preferred pitch of a cholesteric phase, $2\pi/q$. By minimizing the free energy per unit volume of fibril with respect to $\psi(r)$, the equilibrium fibril radius, $R_{\text{eq}}$, and the surface twist angle, $\psi(R_{\text{eq}})$ were determined for different values of the model parameters. These $R_{\text{eq}}$ and $\psi(R_{\text{eq}})$ were then compared with experimental findings. The model \cite{Brown:softmatter2014} showed good agreement with corneal fibrils, which have small radius and large surface tilt. However, it was unclear whether it could also capture the smaller surface tilt and the broad range of radii observed for tendon fibrils.

The physical mechanism of collagen fibril formation {\em in vivo}, as well as the self-assembly of tropocollagen molecules into collagen fibrils {\em in vitro}, is poorly understood. {\em In vitro} studies \cite{Mosser:2006da} have demonstrated that uniform fibril formation will occur without cross-linking or other non-equilibrium processes. This suggests that an equilibrium description of fibrils is appropriate, at least for {\em in vitro} fibrillogenesis. The importance of collagen in biotechnology applications is therefore sufficient motivation for us to further explore the equilibrium picture of radial fibril structure. However, it is attractive to hypothesize that fibrillogenesis {\em in vivo} also exploits equilibrium self-assembly processes.  Better understanding whether and how equilibrium processes could lead to observed radial collagen structures would help us identify when non-equilibrium processes may also be affecting fibril structure. 

In this paper, we use an efficient numerical relaxational method to expand on previous work with the double-twist model, which allows us to map out equilibrium values of fibril radius, surface twist angle, and energy per unit volume of fibril within the entire parameter space of stable fibrils. Using dimensional analysis, we show that just three reduced parameters fully control the experimentally observable behaviour of the system. We use this comprehensive approach to confront both corneal and tendon fibril phenomenology.

%%%%%%%%%%% MODEL %%%%%%%%%
\section{Model}
%%%%%%%%%%%%%%%%%%%%
\subsection{Elastic free energy density}
Individual tropocollagen molecules within collagen fibrils are essentially rod-like, with a length of $\sim\SI{300}{\nano\meter}$, and a diameter of $\sim\SI{1.5}{\nano\meter}$.  To describe the molecular orientation within fibrils, we use a  director field, $\bm{n}(\bm{r})$, which is a unit vector that represents the local, average orientation of molecules within the fibril.   

Following earlier work \cite{Brown:softmatter2014}, we propose that the fibril free energy depends on elastic energy contributions from the orientation field $\bm{n}(\bm{r})$ together with an interfacial energy. We use a leading order gradient expansion for the elastic contributions. The elastic free energy density \cite{Ferrarini:liqcryst2010} of a chiral liquid crystal system with no external stress is 
\begin{align}\label{eq:basic} %%%%%%% equation 1 %%%%%%
f_{\text{el}}=& \frac{1}{2}K_{11}\left(\nabla\cdot\bm{n}\right)^2+\frac{1}{2}K_{22}\left(\bm{n}\cdot\nabla\times\bm{n}+\frac{k_2}{K_{22}}\right)^2\nonumber\\
&+\frac{1}{2}K_{33}(\bm{n}\times\left(\nabla\times\bm{n})\right)^2+k_{13}\nabla\cdot\left(\nabla\cdot\bm{n}\right)\bm{n}\nonumber\\
&-\frac{1}{2}(K_{22}+k_{24})\nabla\cdot\left(\bm{n}\times(\nabla\times\bm{n})+\bm{n}(\nabla\cdot\bm{n})\right),
\end{align}
where we have taken $f_{\text{el}}=0$ in the cholesteric phase. From the last two terms, we see that it is possible to have $f_{\text{el}}<0$ even when all elastic constants are positive.  In using this free energy, we assume that any gradients in $\bm{n}$ are slowly varying compared to the molecular length scale ($\simeq\SI{1.5}{\nano\meter}$). Higher order gradient terms are thereby ignored \cite{Nehring:jchemphys1971}.

Each term in eqn.~\ref{eq:basic} corresponds to a specific distortion. The terms with $K_{11}$, $K_{22}$, and $K_{33}$ correspond to the usual splay, twist, and bend deformations \cite{Sheng:1976bq}, and are always greater than zero. $k_2$ is the ``chiral strength'' and can be of either sign.  $k_{13}$ and $k_{24}$ are the splay-bend and saddle-splay elastic constants, respectively. The terms with $k_{13}$ or $k_{24}$ can be negative, and when integrated will appear as surface terms. They   contribute to equilibrium phases that have a proliferation of interfaces, such as a system of collagen fibrils.

%%%%%%%%%%%%%%%%%%%%%%%%%%%
\subsection{Cholesteric and double-twist  fibril phases}
Equilibrium phases of collagen molecules are determined by the form of $\bm{n}$ that minimizes the total free energy of the system. We consider two phases. The first is a bulk cholesteric phase, which has been observed for concentrated tropocollagen solutions {\em in vitro} \cite{Mosser:2006da}. The director field is e.g. $\bm{n}=\cos(q z)\hat{\bm{x}}+\sin(q z)\hat{\bm{y}}$, where $q \equiv k_2/K_{22}$ here determines the inverse cholesteric pitch of the cholesteric phase. Inserting this into eqn.~\ref{eq:basic} gives $f=0$. Since the cholesteric phase is a bulk phase, any surface effects are negligible and the total  free energy per unit volume, ${E}_{\text{cholesteric}}=0$, for all values of the elastic constants. Any phase with bulk average free energy density ${E}<0$  is therefore thermodynamically stable with respect to the cholesteric phase.

The second phase we consider has individual fibrils with a double-twist  director field \cite{Brown:softmatter2014},
\begin{equation}\label{eq:dtdirector} %%%%%%% equation 2
	\bm{n}=-\sin\psi(r)\hat{\bm{\phi}}+\cos\psi(r)\hat{\bm{z}},
\end{equation}
where $\psi(r)$ is the angle between the director field and the fibril axis. $\psi(R)$ is then the "surface twist" (molecular tilt) of a fibril of radius $R$.  Since we are interested in the radial structure, we ignore contributions from axial packing (e.g. D-banding) of collagen molecules along the fibril. This amounts to an assumption that coupling between radial and axial structure is weak (see Discussion).  Excluding radial/axial coupling greatly simplifies our calculations. 

A surface energy term must be included to account for the cost of creating an interface between individual fibrils and the surrounding fluid. For a single fibril, the free energy per unit length is then
\begin{equation}\label{eq:EperL} %%%%%%% equation 3
E_L \equiv 2\pi\int_0^R rf_{\text{fibril}}(r,\psi(r),\psi'(r))dr+2\pi\gamma R
\end{equation}
where $2\pi\gamma R$ is the energetic cost of the interface between the fibril of cross-sectional circumference $2\pi R$ and its surroundings \cite{Brown:softmatter2014}. (Note that while the bulk $k_{13}$ and $k_{24}$ terms of eqn.~\ref{eq:basic} integrate  mathematically into surface contributions, they are distinct from the interfacial cost $\gamma$.)  The cross-sectional area of a single fibril is $\pi R^2$. Thus,
\begin{equation}\label{eq:EperV} %%%%%%%% equation 4 
	E(R)=\frac{E_L}{\pi R^2} =\frac{2}{R^2}\int_0^Rrf_{\text{fibril}}dr+\frac{2\gamma}{R},
\end{equation}
where $E$ is the total free energy per unit volume of fibril.  We refer to the relationship between $E$ and $R$ as the energy landscape.

Using the double-twist structure eqn.~\ref{eq:dtdirector} in the elastic free energy density eqn.~\ref{eq:basic} gives  the free energy density \cite{Brown:softmatter2014},
\begin{align}\label{eq:ffibril} %%%%%%% equation 5
f_{\text{fibril}}=&\frac{1}{2}K_{22}\left(q-\psi'-\frac{\sin2\psi}{2r}\right)^2+\frac{1}{2}K_{33}\frac{\sin^4\psi}{r^2}\nonumber\\
&-\frac{1}{2}(K_{22}+k_{24})\frac{1}{r}\frac{d\sin^2\psi}{dr}.
\end{align}
where here $q=k_2/K_{22}$ is the chiral wavenumber of the double-twist phase. Note that the $K_{11}$ and $k_{13}$ terms have dropped out since $\nabla\cdot\bm{n}=0$ for double-twist.

Minimizing eqn.~\ref{eq:EperV} with respect to the function $\psi(r)$ using standard calculus of variations techniques \cite{Brown:softmatter2014}, we arrive at the boundary value problem 
\begin{subequations}\label{eqs:BVP} %%%%%%% equation 6
\begin{align}
(r\psi')'&=q+\frac{K_{33}}{K_{22}}\frac{\sin(2\psi)}{r}\sin^2\psi-\cos(2\psi)\left(q-\frac{\sin(2\psi)}{2r}\right),\label{eq:ODE}\\
\psi(0)&=0,\label{eq:psi0}\\
\psi'(R)&=q+\frac{k_{24}}{K_{22}}\frac{\sin(2\psi(R))}{2R},\label{eq:psiR}
\end{align}
\end{subequations}
where eqn.~\ref{eq:psiR} is a natural boundary condition which follows from the functional minimization procedure, and $\psi' \equiv d\psi/dr$. We must have $\psi(0)=0$, as any non-zero twist at $r=0$ would imply singular $f_{\text{fibril}}$ and an infinite $E$ from eqns.~\ref{eq:EperV} and \ref{eq:ffibril}.

\subsubsection{Dimensional Analysis} %%%%%%%
While there are five parameters which control the behavior of our model, $q$, $\gamma$, $K_{22}$, $K_{33}$, and $k_{24}$, we can reduce this to three dimensionless variables (see Appendix~\ref{dimensionless}), $\tilde{K}_{33}=K_{33}/K_{22}$, $\tilde{\gamma}=\gamma/(K_{22}q)$, and $\tilde{k}_{24}=k_{24}/K_{22}$, which we utilize for the remainder of the paper. This lets us express quantities of interest in terms of dimensionless parameter combinations: 
\begin{subequations}\label{eq:dimensional} %%%%%% equation 7
\begin{align}
\tilde{E}&=g_1\left(qR,\tilde{K}_{33},\tilde{\gamma},\tilde{k}_{24}\right), \label{eq:Edimension}\\
qR&=g_2\left(\tilde{K}_{33},\tilde{\gamma},\tilde{k}_{24}\right), \label{eq:Rdimension}\\
\psi(qr)&=g_3\left(qr,\tilde{K}_{33},\tilde{\gamma},\tilde{k}_{24}\right), \label{eq:psidimension}
\end{align}
\end{subequations}
where the functions $g_1$, $g_2$, and $g_3$ are determined numerically, $\tilde{E}\equiv E/(K_{22}q^2)$, and we solve $\psi$ as a function of dimensionless radius $q r$.  We have reduced our parameter space from five to three dimensions, together with an inverse length $q$ that sets the scale for $R$.

The elastic constants for collagen solutions are not well documented. We use values determined experimentally from liquid crystal systems with molecules similar to tropocollagen molecules. For the polypeptide $\alpha$-helical chain poly-$\gamma$-benzyl-L-glutamate (PBLG), the ratio of bend to twist elastic constant saturates at $K_{33} \simeq 30K_{22}$ for aspect ratios $L/D \gtrsim100$, where $L$ is the length and the diameter $D$ of PBLG is between $\SI{1.5}{\nano\meter}$ to $\SI{2.5}{\nano\meter}$ \cite{Lee:liqcryst1990}.  The aspect ratio of tropocollagen, $L/D=200$, then leads us to use $\tilde{K}_{33}=30$ for this paper. (In Appendix~\ref{K33} we explore the effects of different $\tilde{K}_{33}$ values on our results for the surface twist.) Differences in solution conditions, molecular composition, and concentration can in principle affect $\tilde{K}_{33}$ \cite{Onsager:1949jk, Straley:1976, Tortora:2017}, however approximately the same ratio is observed over a range of temperature and concentration in long-aggregates of lyotropic chromonic liquid crystals \cite{Zhou:2014dk}. 

%%%%%%%%%%%%%%%%%
\subsection{Energy Minimization}
We solve eqns~\ref{eqs:BVP} numerically using finite-difference relaxation  \cite{Press:1992td}. We have also derived an explicit (but unwieldy) power-series solution, see Appendix~\ref{power-series}. We use the leading cubic terms of this power-series as an initial guess for our relaxation approach, and use higher-order solutions as occasional checks that the relaxation approach has converged. The iterated relaxation converges on the $\psi(qr)$ that minimizes the dimensionless version of eqn.~\ref{eq:EperV} for a selected $qR$.  We repeat this procedure for different $qR$ to determine the energy landscape, $\tilde{E}(qR)$, for a given parameter set \cite{Brown:softmatter2014}.

We are particularly interested in the dimensionless radius $qR_{\text{eq}}$ that minimizes $\tilde{E}(qR)$. To find $qR_{\text{eq}}$, we used a standard golden ratio search. Our search bounds were $qR\in\left[\SI{e-5}{},1\right]$. If $\tilde{E}(qR_{\text{eq}})\equiv\tilde{E}_{\text{eq}}<0$ for a set of parameter values, then the bulk fibril phase is an equilibrium phase with respect to the cholesteric for those parameters. To avoid cumbersome notation, we will use the equilibrium result $\psi(qr)\equiv\psi_{\text{eq}}(qr)$ unless otherwise noted.

%%%%%%%% RESULTS %%%%%%%%%%%
\section{Results}
%%%%%%%%%%%%%%%%%%%%
\subsection{Narrow equilibrium regime}
\begin{figure}[!htb]
\hspace{-10pt}
	\includegraphics[width=0.5 \textwidth]{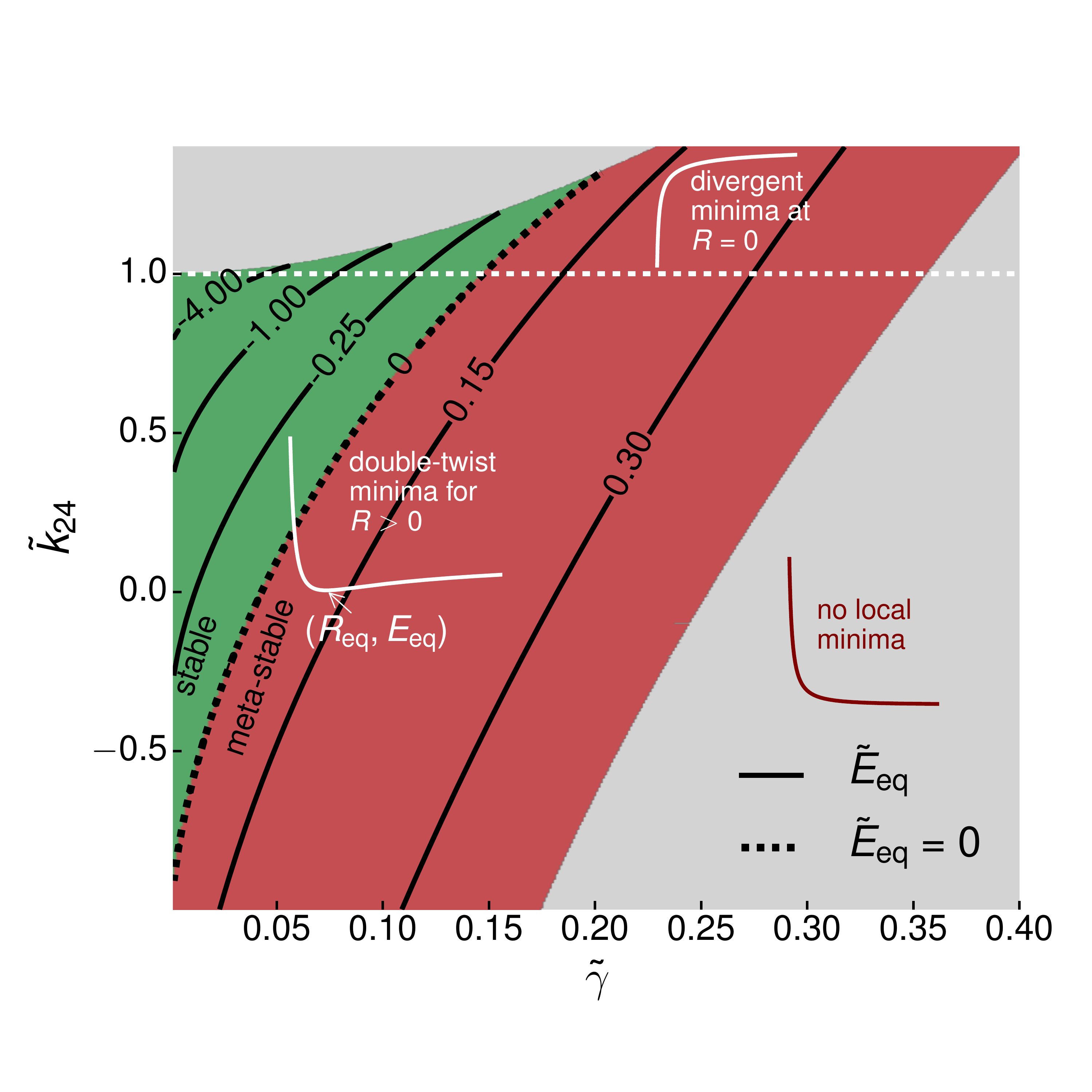}
\caption{The green and red regions in this $\tilde{k}_{24}$ vs $\tilde{\gamma}$ (with $\tilde{K}_{33}=30$) phase plot indicate possible fibril phases. The green region indicates the existence of double-twist fibrils that are stable with respect to the cholesteric phase, with $E_{\text{eq}}<0$. The red region indicates meta-stable minima with respect to the cholesteric phase, with $E_{\text{eq}}\geq0$. Contours indicate the values of the dimensionless free energy density $\tilde{E}_{\text{eq}}\equiv E_{\text{eq}}/(K_{22} q^2)$. The inset white curve labeled "double-twist minima for $R>0$" demonstrates a typical relationship between $E$ and $R$ for values of $\tilde{\gamma}$ with $\tilde{k}_{24}\leq1$. For $\tilde{k}_{24}>1$ (above dashed white line), there is an additional, divergent global minimum as $R \rightarrow 0$, illustrated by the inset curve labeled "divergent minima at $R=0$". Gray regions do not have any local minima with $0<R<\infty$. Note that $\tilde{k}_{24}\equiv k_{24}/K_{22}$, $\tilde{\gamma}\equiv\gamma/(K_{22} q)$, and $\tilde{K}_{33}\equiv K_{33}/K_{22}$.}\label{fig:EnergyLandscape}
\end{figure}

In Fig.~\ref{fig:EnergyLandscape}, we show the \textit{global} energy landscape for double-twist collagen fibrils as the dimensionless parameters $\tilde{\gamma}$ and $\tilde{k}_{24}$ are varied. The ratio $\tilde{K}_{33}=30$ is held constant. The colour and contours represent the dimensionless  minimum energy, $\tilde{E}_{\text{eq}}$, for double-twist fibrils --- green  indicates equilibrium fibrils with respect to the cholesteric phase, while red indicates metastable fibrils. We see that there is only a small region of equilibrium fibrils, where we require $-1 \leq \tilde{k}_{24} \lesssim 1.2$ and $\tilde{\gamma} \lesssim 0.2$.  The minimum fibril energy $\tilde{E}_{\text{eq}}$ increases monotonically with increased $\tilde{\gamma}$ or with decreasing $\tilde{k}_{24}$.  

%%%%%%%%%%%% FIGURE 1 %%%%%%%%%%%%%%%%%

The energies shown in Fig.~\ref{fig:EnergyLandscape} represents the energy of double-twist fibrils that have a finite radius $R$. For larger values of $\tilde{\gamma}$ there is no local minimum at  $R>0$ (gray region), and we would instead expect to observe a bulk cholesteric phase. This is also what we expect in most of the metastable regime, and arises because the energy cost of the interface in a fibril phase is large due to the surface tension $\gamma$.  

For $\tilde{k}_{24}>1$, we observe a divergent minimum energy for $R \rightarrow 0$. When both a divergent minimum for $R \rightarrow 0$ and a local minimum at finite $R$ is present, we illustrate the local minimum behaviour only (i.e. shading and contours in non-gray regions with $\tilde{k}_{24}>1$ represent the local minima). This divergent minimum arises because sufficiently large $k_{24}$ encourages interface proliferation in the fibril phase. This can be seen explicitly with eqns.~\ref{eq:EperV} and \ref{eq:ffibril} using a linearly varying ansatz for the pitch, $\psi = r \psi(R)/R$. For $\psi(R) \ll 1$ we obtain $E = \psi(R)^2 (K_{22}-k_{24})/R^2 + 2 \gamma/R$. For $k_{24}>K_{22}$ we obtain $E \rightarrow -\infty$ as $R \rightarrow 0$. However, this singular solution is for a continuum model where  fibril radii are large with respect to the diameter of individual molecules, $d=\SI{1.5}{\nano\meter}$. We would also expect higher order gradient terms, absent in  eqn.~\ref{eq:basic}, to  change (and perhaps eliminate) the singular solution at $R \approx 0$ for $\tilde{k}_{24}>1$. 

To confront our double-twist solutions with experimental measurements of collagen fibrils, we investigate our model's predictions of surface twist, $\psi_{R_{eq}}\equiv\psi(qR_{\text{eq}})$, and fibril radius, $R_{\text{eq}}$.

%%%%%%%%%%%%%%%%%%%%%%%%%%
\subsection{Experimental observables: Surface twist and fibril radius}
%%%%%%%%% FIGURE 2 (SURFACE TWIST) %%%%%%%%%
\begin{figure}[!tbh]
\hspace{-10pt}
\includegraphics[width=0.5\textwidth]{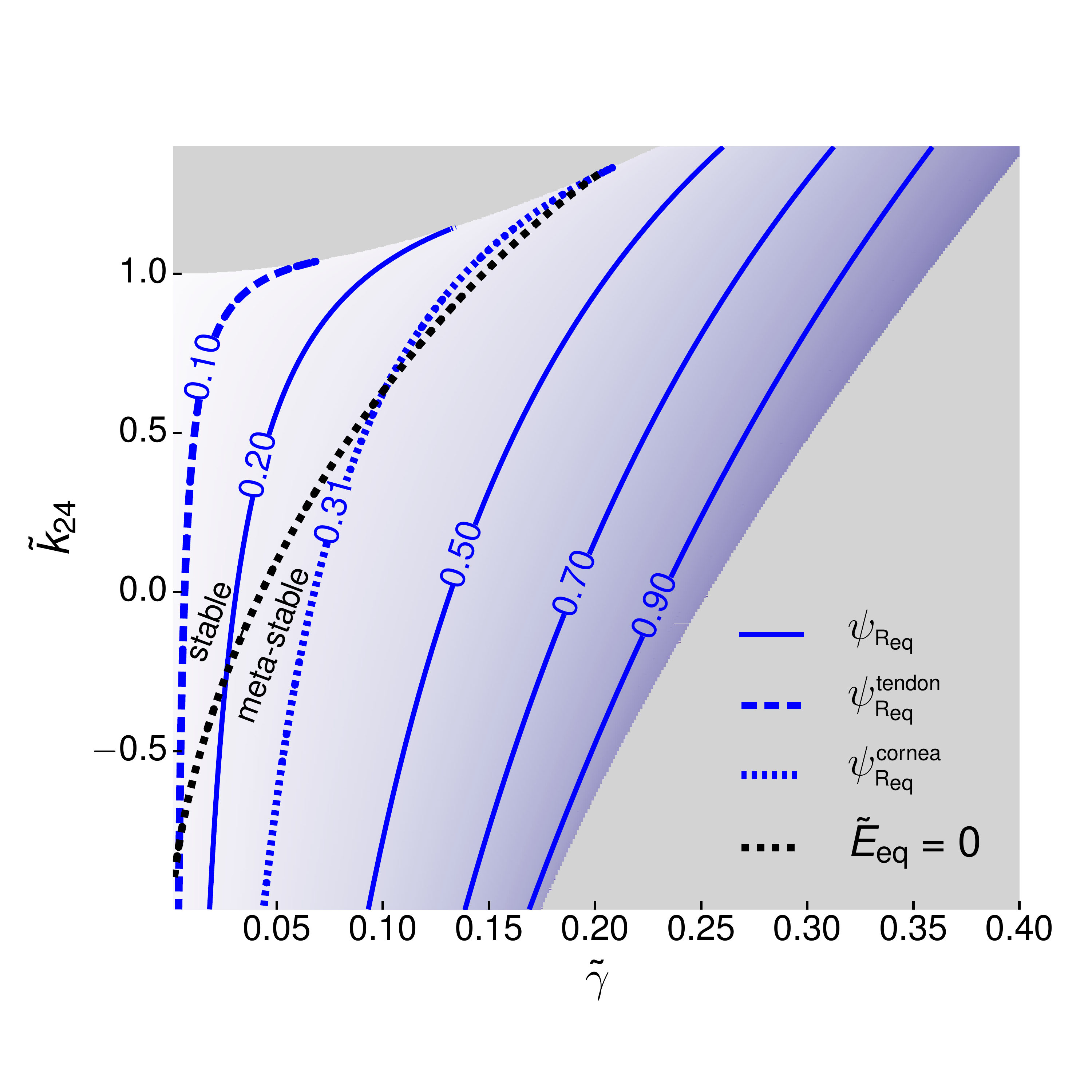}
\caption{Contours of surface twist $\psi_{R_{\text{eq}}}\equiv\psi(qR_{\text{eq}})$ (solid and dashed blue) in radians, vs the reduced saddle-splay elastic constant $\tilde{k}_{24}$ and the reduced surface tension $\tilde{\gamma}$, all with $\tilde{K}_{33}=30$. $\psi_{R_{\text{eq}}}=\SI{0.1}{\radian}$ and $\psi_{R_{\text{eq}}}=\SI{0.31}{\radian}$ are typical surface twists observed in tendon and cornea fibrils, respectively, and are distinguished above with dashed contour lines. Meta-stable ($E_{\text{eq}}\geq0$) and stable ($E_{\text{eq}}<0$) fibril phases with respect to the bulk cholesteric phase are separated by the black, dashed line. The gray areas correspond to parameter space regions for which no stable or meta-stable double-twist configurations are found. Note that $\tilde{k}_{24}\equiv k_{24}/K_{22}$, $\tilde{\gamma}\equiv\gamma/(K_{22} q)$, and $\tilde{K}_{33}\equiv K_{33}/K_{22}$.}\label{fig:PsiRlandscape}
\end{figure}

Fig.~\ref{fig:PsiRlandscape} shows the surface twist landscape. Corresponding with Fig.~\ref{fig:EnergyLandscape}, the gray regions at the upper left and to the right have no fibril phases.  $\psi_{R_{\text{eq}}}$ increases with increasing $\tilde{\gamma}$ and decreasing $\tilde{k}_{24}$, with blue lines of constant $\psi_{R_{\text{eq}}}$ (in radians) shown.  Double-twist phases that are stable with respect to the bulk cholesteric phase occur to the left of the black dashed line ($E_{\text{eq}}<0$), as indicated. 

Two surface twist values of particular interest are $\psi_{R_{\text{eq}}}=\SI{0.1}{\radian}$ and $\psi_{R_{\text{eq}}}=\SI{0.31}{\radian}$, being typical surface twist angles observed in tendon fibril and corneal fibril, respectively. We have labeled these two values of surface twist with blue dashed lines in Fig.~\ref{fig:PsiRlandscape}. Furthermore, other types of fibrils in vivo tend to have smaller surface twists than corneal fibrils $\leq\SI{0.31}{\radian}$, which gives the corneal dashed line in Fig. ~\ref{fig:PsiRlandscape} further meaning as an upper limit of surface twist values observed in vivo \cite{Reale:1981, Ruggeri:1979hw, Raspanti:interjbiomacromol1989}. Remarkably, this upper bound of surface twist approximately coincides with the \textit{stable equilibrium} regime of double-twist fibrils (i.e. the region to the left of the black dashed line in Fig. ~\ref{fig:PsiRlandscape}).  

%%%%%%%% FIGURE 3 (FIBRIL RADIUS) %%%%%%%%%%%
\begin{figure}[!tbh]
\hspace{-10pt}
\includegraphics[width=0.5\textwidth]{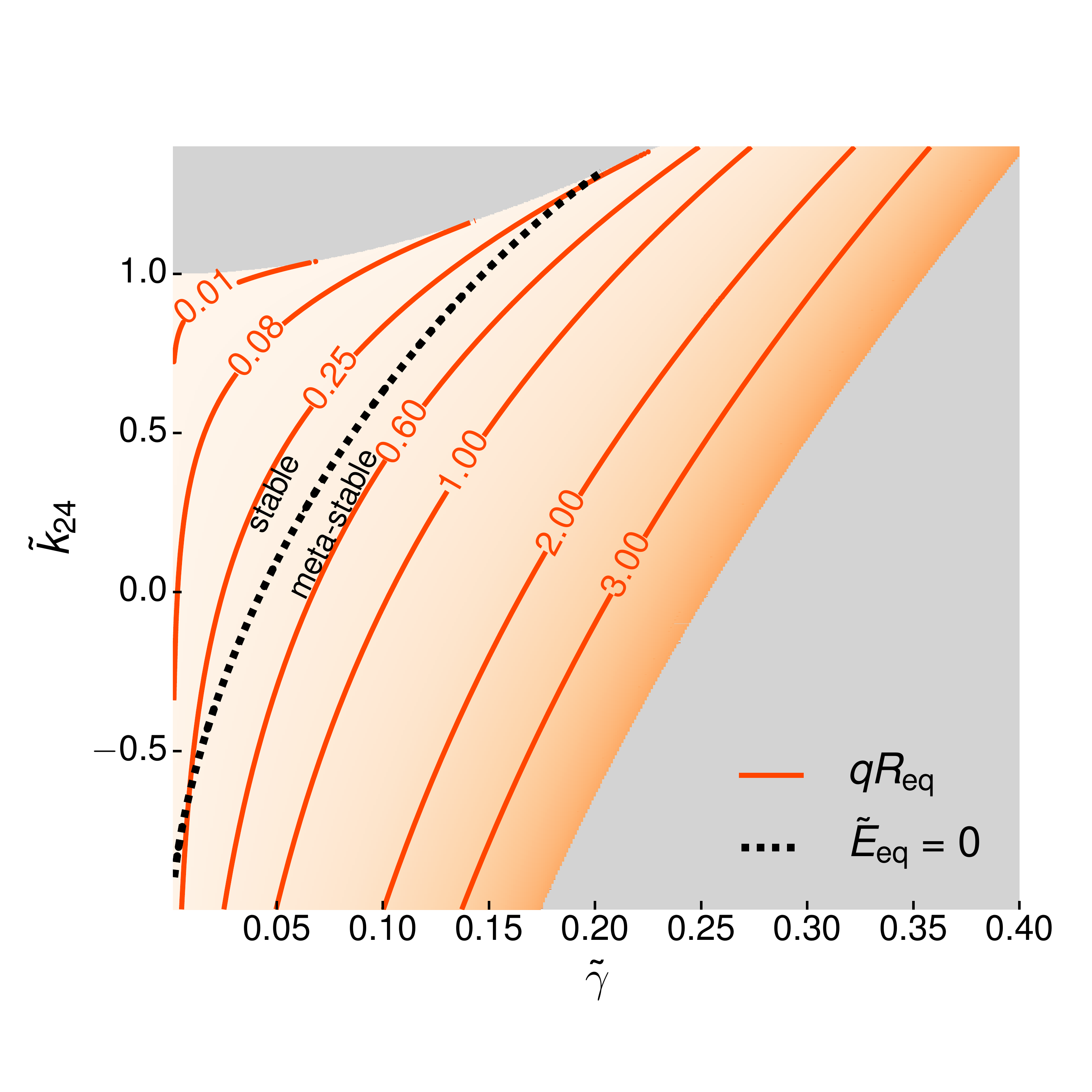}%\vspace{-30pt}
\caption{Contours of scaled equilibrium fibril radius $q R_{\text{eq}}$  as a function of the reduced saddle-splay elastic constant, $\tilde{k}_{24}$, and the dimensionless surface-tension $\tilde{\gamma}$, all with $\tilde{K}_{33}=30$. $q R_{\text{eq}}$ increases with increasing $\tilde{\gamma}$, and decreases with increasing $\tilde{k}_{24}$. Values of $q R_{\text{eq}}$ to the left of the black, dashed line are stable with respect to the bulk cholesteric phase ($\tilde{E}_{\text{eq}}<0$). The gray areas correspond to parameter space regions for which no stable or meta-stable double-twist configurations are found. Note that $\tilde{k}_{24}\equiv k_{24}/K_{22}$, $\tilde{\gamma}\equiv\gamma/(K_{22} q)$, and $\tilde{K}_{33}\equiv K_{33}/K_{22}$.}
\label{fig:Rqlandscape}
\end{figure}

We also obtain reduced equilibrium fibril radii, $q R_{\text{eq}}$, as shown in  Fig.~\ref{fig:Rqlandscape}.  As a consequence of eqn.~\ref{eq:Rdimension}, we do not obtain the radii directly.  We see that $q R_{\text{eq}}$ increases with increasing $\tilde{\gamma}$, and decreases with increasing $\tilde{k}_{24}$ --- the same qualitative behavior as $\psi_{R_{\text{eq}}}$.  For fixed $q$ the behavior of $R_{\text{eq}}$ as other parameters are varied is immediately given: the radius decreases as $k_{24}$ increases, or as the surface tension $\gamma$ decreases. Increasing $K_{22}$ moves directly towards the origin, and can either increase $R_{\text{eq}}$ (for fibrils with small $\psi_{R_{\text{eq}}}$) or decrease $R_{\text{eq}}$ (for fibrils with $\psi_{R_{\text{eq}}} \gtrsim \SI{0.2}{\radian}$, or $\SI{10}{\degree}$). If we increase $q$ and leave other parameters fixed, we see that the scaled surface-tension $\tilde{\gamma}$ will decrease --- leading to smaller $q R_{\text{eq}}$ values. Since we have increased $q$, we then obtain even smaller $R_{\text{eq}}$ values.  

%%%%%%%%% FIGURE 4 (solution at selected points ) %%%%%%%%%%%
\begin{figure*}[!tbh]
	\centering
	\includegraphics[width=\textwidth]{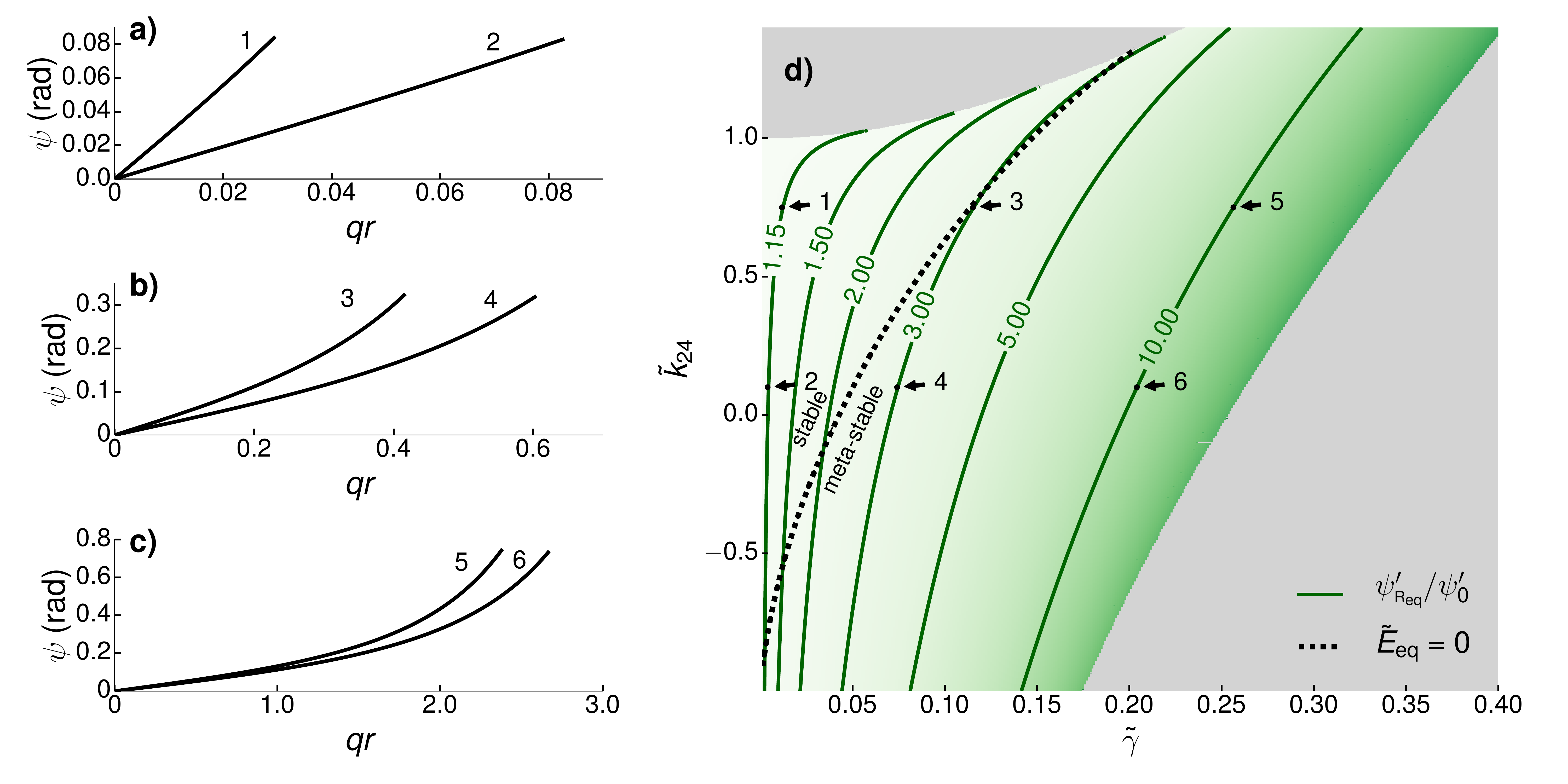}
    {\phantomsubcaption\label{fig:LowNonlinearity}}
	{\phantomsubcaption\label{fig:MidNonlinearity}}
	{\phantomsubcaption\label{fig:HighNonlinearity}}
	{\phantomsubcaption\label{fig:NonlinearityGrid}}
	\captionsetup{aboveskip = -10pt}
	\caption{In \protect\subref{fig:LowNonlinearity}-\protect\subref{fig:HighNonlinearity}, six different double-twist configurations, $\psi$ vs scaled radial distance $qr$, are illustrated for the parameter values indicated in \protect\subref{fig:NonlinearityGrid} -- with corresponding labels from 1-6. 1, 3 and 5 are points on the $\tilde{k}_{24}=0.75$ line; 2, 4, and 6 are points on the $\tilde{k}_{24}=0.1$ line. Both $\psi$ and $\psi'$ increase monotonically with $r$ for all parameter values. The contours in  \protect\subref{fig:NonlinearityGrid} indicate the ratio of surface twist gradient at the surface to that in the fibril centre, $\psi'(qR_{\text{eq}})/\psi'(0)\equiv\psi'_{R_{\text{eq}}}/\psi'_0$, which captures non-linearities in the double-twist configuration. As before, the black dashed line separates fibrils that are stable with respect to the bulk cholesteric phase (left of line) from those which are only meta-stable (right of line).  The gray areas of \protect\subref{fig:NonlinearityGrid} correspond to parameter space regions for which no stable or meta-stable double-twist configurations are found.}\label{fig:Nonlinearity}
\end{figure*}
%here: figure 4

%%%%%%%%%%%%%%%%%%%%%%%%%%%%
\subsection{Non-linearity of twisting within fibril}
From our free energy functional, at the fibril centre collagen molecules are aligned with the fibril axis, with $\psi(0)=0$.  For $r>0$,  we illustrate $\psi(qr)$ for six parameter values in Fig.~\ref{fig:LowNonlinearity}-\subref{fig:HighNonlinearity}.  All of the curves exhibit two properties: 1) $\psi(qr)$ increases monotonically with $qr$, and 2) the twist gradient also increases with radius, i.e. $\psi''(qr)>0$. With these two properties in mind, we quantify the double-twist nonlinearity with the ratio of the twist angle gradient at the fibril surface, $\psi'_{R_{\text{eq}}}\equiv\psi'(qR_{\text{eq}})$, to the twist angle gradient at the fibril centre, $\psi'_0\equiv\psi'(0)$, as shown in Fig.~\ref{fig:NonlinearityGrid}. Nonlinearity increases with increasing $\tilde{\gamma}$, and decreases with increasing $\tilde{k}_{24}$. We see that equilibrium fibrils may have significant twist nonlinearities, up to $\psi'_{R_{\text{eq}}}/\psi'_0 \approx 3$.

%%%%%%%%%%% DISCUSSION %%%%%%%%%%%
\section{Discussion}
We identified dimensionless parameter combinations (eqn.~\ref{eq:dimensional}) that reduced the number of independent parameters in our equilibrium free energy density (eqn.~\ref{eq:EperV}) for the collagen orientation within double-twist fibrils  (eqn.~\ref{eq:dtdirector}). We solved the dimensionless equations numerically, and identified a narrow parameter regime (green region of Fig.~\ref{fig:EnergyLandscape}) that produces double-twist fibrils that are thermodynamically stable with respect to a bulk cholesteric phase. 

The parameters of our model are the coarse-grained elastic constants that determine the free energy costs of spatial-gradients of the collagen orientation ($K_{22}$, $K_{33}$, $K_{22}$, $k_{24}$) together with a surface energy $\gamma$ and a chiral wavenumber $q$. One dimensionless parameter combination is relatively well determined by the long semi-flexible configuration of individual collagen molecules  ($K_{33}/K_{22}=30$). Remarkably,  we find that only two dimensionless parameter combinations ($k_{24}/K_{22}$ and $\gamma/(K_{22} q)$) are then required to determine both the  surface twist $\psi_{R_{\text{eq}}}$ (Fig.~\ref{fig:PsiRlandscape}) and the dimensionless  radius $qR_{\text{eq}}$ (Fig.~\ref{fig:Rqlandscape}) of equilibrium collagen fibrils. 

We find that equilibrium surface twists should all satisfy an upper bound: $\psi_{R_{\text{eq}}} \leq \SI{0.33}{\radian}$ ($\SI{19}{\degree}$), which approximately coincides with the maximum surface twist reported in the {in vivo}  literature \cite{Holmes:2001bt}.  

%%%%%%% PRESCRIPTIVE %%%%%%%%%%%%%%%%%%%%
\subsection{Polymorphism of  collagen fibrils}
A surprise in considering Figs.~\ref{fig:EnergyLandscape}, \ref{fig:PsiRlandscape}, and \ref{fig:Rqlandscape} is the wide range of equilibrium configurations available to collagen fibrils over a relatively narrow parameter regime. This polymorphism allows different aspects of fibril structure to be emphasized for different parameterizations. 

\subsubsection{Collagen fibril stability}  %%%%%%%%%%%%%%
The thermodynamic stability with respect to the cholesteric phase is assessed by the free energy per unit volume, as illustrated in Fig.~\ref{fig:EnergyLandscape}. We see that the most stable (lowest energy) fibrils are in the upper-left corner with a combination of small $\gamma$ and large $K_{22}$ and $q$ --- above point ``1'' in Fig.~\ref{fig:Nonlinearity}, with $k_{24} \simeq K_{22}$. Note that what is presented is $E_{\text{eq}}/(K_{22} q^2)$, so that with large $K_{22}$ and $q$ the cohesion energy is even larger. 

One consequence of selecting for more stable fibrils is that the expected surface twist values would be quite small, according to Fig. ~\ref{fig:PsiRlandscape}. Interestingly, we would expect a uniform twist gradient (Fig.~\ref{fig:Nonlinearity}) in this regime as well. In contrast, to allow for fibrils with larger surface twist, $\tilde{\gamma}$ must be fine-tuned to values near the stability boundary --- close to point ``3'' in Fig.~\ref{fig:Nonlinearity} --- making fibrils with large surface twist (and nonlinear $\psi(qr)$) less thermodynamically stable than their small twist, linear counterparts. We note that all fibrils which are stable with respect to the cholesteric have $\psi_{R_{\text{eq}}}\leq\SI{0.33}{\radian}$. 

The relationship between thermal stability and fibril radius is complicated by the scaling of $R_{\text{eq}}$ with $q$, as the contours in Fig.~\ref{fig:Rqlandscape} depend on $q$ as well as $\tilde{\gamma}$ and $\tilde{k}_{24}$. Thus, to investigate the relationship between thermal stability and fibril size, we look at the two ways in which large (small) radius equilibrium fibrils can be generated from our model. The first is to maximize (minimize) $qR_{\text{eq}}$ at a constant $q$. From Fig.~\ref{fig:Rqlandscape}, this would be achieved by fine-tuning $\tilde{\gamma}$ close to (far from) the stability boundary. This approach would indicate that smaller fibrils are more thermodynamically stable than large fibrils.

The second approach to generate large (small) fibrils is to decrease (increase) the chiral wavenumber $q$ at a constant $qR_{\text{eq}}$, while also keeping $\tilde{\gamma}$ and $\tilde{k}_{24}$ constant. In this approach, you would stay at the same point in Fig. ~\ref{fig:EnergyLandscape} and ~\ref{fig:Rqlandscape}, and so $E_{\text{eq}}/(K_{22}q^2)$ and $qR_{\text{eq}}$ would remain constant. As you decrease (increase) $q$, fibril radius increases (decreases), but thermodynamic stability decreases (increases) as well. Thus, both approaches to increasing fibril radius tend to decrease thermal stability. Given this prediction, it is unclear what functional role large fibrils might have, if it is not to increase stability. While large fibrils are expected to be individually stronger than small ones, the packing fraction of large or small fibrils would be the same and so would bulk moduli of closely packed fibrils.

\subsubsection{Influence of Collagen types}%%%%%%%%%%%%%%%
Collagen fibrils \textit{in vivo} generally contain a tissue-dependent mixture of collagen types \cite{Wess:2005, Raspanti:2018}. For example,  while well-studied tendon and corneal fibrils are predominantly composed of type-I collagen they contain an admixture of type-III collagen \cite{Keene:1987}.   The best characterized heterotypic mixtures \textit{in vitro} has been blends of types I and III collagen \cite{Fleischmajer:1990, Geerts:1990, Asgari:2017}, though I/V \cite{Birk:1990, Adachi:1986} and II/III blends \cite{Young:2000} have also been studied. 

The distribution of collagen types within individual fibrils has been qualitatively assessed from immunoassay double-labeling. Both type I and type III are seen on fibril surfaces \cite{Geerts:1990, Fleischmajer:1990, Asgari:2017} indicative of homogeneity (the evidence is, however, mixed\cite{Raspanti:2018}). Under the assumption that mixtures of collagen types are spatially homogeneous within a fibril, the elastic parameters of the mixture should be interpolations between those of the pure collagen types \cite{Straley:1976}. In which case, our equilibrium picture would apply to heterotypic fibrils --- and the reduced elastic parameters of mixtures would sit on curves between those of the pure types. 

Varying the composition of heterotypic I/III fibrils  leads to variations of fibril radius \cite{Asgari:2017} --- from $\SI{0.1}{\micro\meter}$ (entirely type I) to $\SI{0.025}{\micro\meter}$ (entirely type III).  Our model can reproduce that either by moving the reduced parameters, e.g. $\tilde{\gamma}$, or by changing $q$. Changes to $\tilde{\gamma}$ would be associated with a change in the surface twist, while changes to $q$ could be assessed in the cholesteric phase. However, neither surface twist nor cholesteric $q$ have been systematically characterized in type I/III mixtures. 

\subsubsection{D-band spacing}
While we have assumed that the radial and longitudinal structures are decoupled, a simple projective-coupling has been proposed in the literature \cite{Itoh:1996, Raspanti:2018}, corresponding to the D-band period being reduced by a factor of $\cos(\psi)$ due to non-zero twist.  For our nonlinear double-twist model, the question immediately arises about how a single D-band spacing can represent a continuously varying twist, $\psi(qr)$. We  hypothesize that surface measurements of the D-band period via scanning electron microscopy or atomic force microscopy would probe surface twist $\psi(qR)$ while bulk measurements of the D-band period via transmission electron microscopy or X-ray scattering  would probe a volume-average twist $\langle \cos(\psi(qr)) \rangle$.  Combining both types of measurements on the same set of fibrils would then  provide additional insight into the nature of the radial and longitudinal coupling.

Our model has a maximal surface twist of $\SI{0.33}{\radian}$, and a minimal twist of $0.002\simeq0\:\si{\radian}$, corresponding to at most a $5\%$ difference of D-band spacing between fibrils according to the projective-coupling hypothesis. While surface twist of heterotypic I/III fibrils has not been characterized, the D-band spacing has been \cite{Asgari:2017}.  For $100\%$ collagen-III (compared to pure collagen-I fibrils) there is a significant $39\%$ decrease in the D-band spacing. This exceeds our maximal surface twist effect, but could be attributed to changes in the gap-spacing of the D-band \cite{Antipova:2010} or to rope-like ultrastructure \cite{Bozec:2007} rather than to molecular tilt. Experimentally relating surface twist measurements of fibrils to a more detailed assessment of longitudinal structure and ultrastructure would be desirable to untangle these effects.

%%%%%%%%%%%%%%%%
\subsection{Experimental guidance on elastic parameters}
\label{parameters}
The chiral wavenumber $q$ can be directly assessed  within cholesteric phases through the cholesteric pitch $P=2\pi/q$.  Polarized light microscopy observations of rat tail tendon tropocollagen solubilized in acid show that cholesteric phases emerge at concentrations above $\SI{50}{\milli\gram\per\milli\liter}$, with decreasing pitch from $P\simeq\SI{20}{\micro\meter}$ at $\sim\SI{50}{\milli\gram\per\milli\liter}$ to $P\simeq\SI{0.5}{\micro\meter}$ at $\sim\SI{400}{\milli\gram\per\milli\liter}$ \cite{DeSaPeixoto:2011jm}. While we might expect variation in $q$ for fibrils due to variable solution conditions \cite{Onsager:1949jk, Straley:1976}, we expect a similar range of values $q\in\left[0.1\pi\si{\per\micro\meter}, 4\pi\si{\per\micro\meter}\right]$.

The surface tension, $\gamma$, quantifies the cost of an interface between two bulk phases. In our case, the interface is between individual fibrils and the surrounding aqueous collagen solution. No experimental measurements of $\gamma$ have been reported for collagen.  However, we assume surface-tensions are similar in magnitude to the nematic-isotropic interface for liquid crystal systems. A lower bound of surface tension of an isotropic-nematic interface is that of p-azoxyphenetole, for which  $\gamma \gtrsim\SI{.5}{\pico\newton\per\micro\meter}$ \cite{Kahlweit:1973gr}. Conversely, a larger value of $\gamma$ reported in this type of system is that of MBBA, with $\gamma=\SI{24}{\pico\newton\per\micro\meter}$ \cite{Langevin:1973fk,Chen:1996cv}. Other experimental values fall within this range \cite{Oswald:2015cj, Faetti:1984bn, Williams:1976bd}. Using Onsager's theory of hard rods \cite{Onsager:1949jk}, a theoretical expression of $\gamma$ has been derived for isotropic-nematic interfaces near the phase transition \cite{Doi:1985tv}. Applying this result to our system, we obtain $\gamma\sim\SI{2.3}{\pico\newton\per\micro\meter}$ which is consistent with the experimental bounds. Accordingly, we expect $\gamma\in\left[\SI{.5}{\pico\newton\per\micro\meter}, \SI{25}{\pico\newton\per\micro\meter}\right]$. 

To determine the value of the twist elastic constant, $K_{22}$, for collagen fibrils, we again use typical values of liquid crystal systems. For PBLG, a range of $K_{22}$ values from $\SI{0.6}{\pico\newton}$ to $\SI{6.2}{\pico\newton}$ \cite{DuPre:1975el, Taratuta:1988jo, Toriumi:1998fi} have been measured depending on the solvent used. In these measurements, no significant concentration \cite{Taratuta:1988jo} or molecular weight \cite{Toriumi:1998fi} dependence has been observed. We therefore expect $K_{22}\in\left[\SI{0.6}{\pico\newton}, \SI{6}{\pico\newton}\right]$.

Experimentally determining the saddle-splay elastic constant, $k_{24}$, is difficult due to the surface-like nature that it represents in the free energy. The saddle-splay to twist ratio has been estimated to be $k_{24}/K_{22}\simeq 2$ for nematic systems using deuterium nuclear-magnetic-resonance \cite{Allender:prl1991} and polarization microscopy \cite{Polak:1994iu}. No measurements of $k_{24}$ for long, chiral molecules similar to tropocollagen have been reported. Theoretical calculations predict that $k_{24}=\frac{1}{2}(K_{11}-K_{22})$ \cite{Nehring:jchemphys1971}, which with $K_{11}>K_{22}$ \cite{Kroger:2007, Tortora:2017} implies $k_{24}\geq0$. However, this result was derived through an interaction energy (vs a free energy), and thus is likely valid only for thermotropic systems.

\subsection{Comparison with {\em in vivo} fibril ultrastructure}%%%%%%%%%%%%%%%%
Our theoretical equilibrium treatment highlights the importance of surface twist, since it significantly constrains our model parameterization. (The comparison between experiment and our model is not as definitive when looking at $R_{\text{eq}}$, because we can only constrain the product $q R_{\text{eq}}$.) The surface twist angle measured {\em in vivo} is correlated to the anatomical location of the fibril, as well as the type of tropocollagen found within the fibril \cite{Reale:1981, Ottani:2001, Ottani:micron2002,Raspanti:interjbiomacromol1989, Raspanti:2018}. Two well-studied fibril types {\em in vivo} are corneal fibrils, which have large surface twists $\simeq \SI{0.31}{\radian}$ \cite{Holmes:2001bt}, and tendon fibrils, which have fairly small surface twists $\simeq \SI{0.1}{\radian}$ \cite{Hulmes:1981cy}.  

%%%%%%%%  Corneal fibrils %%%%%%%%%%%
\begin{figure}[!t]
	\hspace{-10pt}
	{\phantomsubcaption\label{fig:E_k24_psiRis0.31}}
	{\phantomsubcaption\label{fig:qR_psiRis0.31}}
	\includegraphics[width=0.5\textwidth]{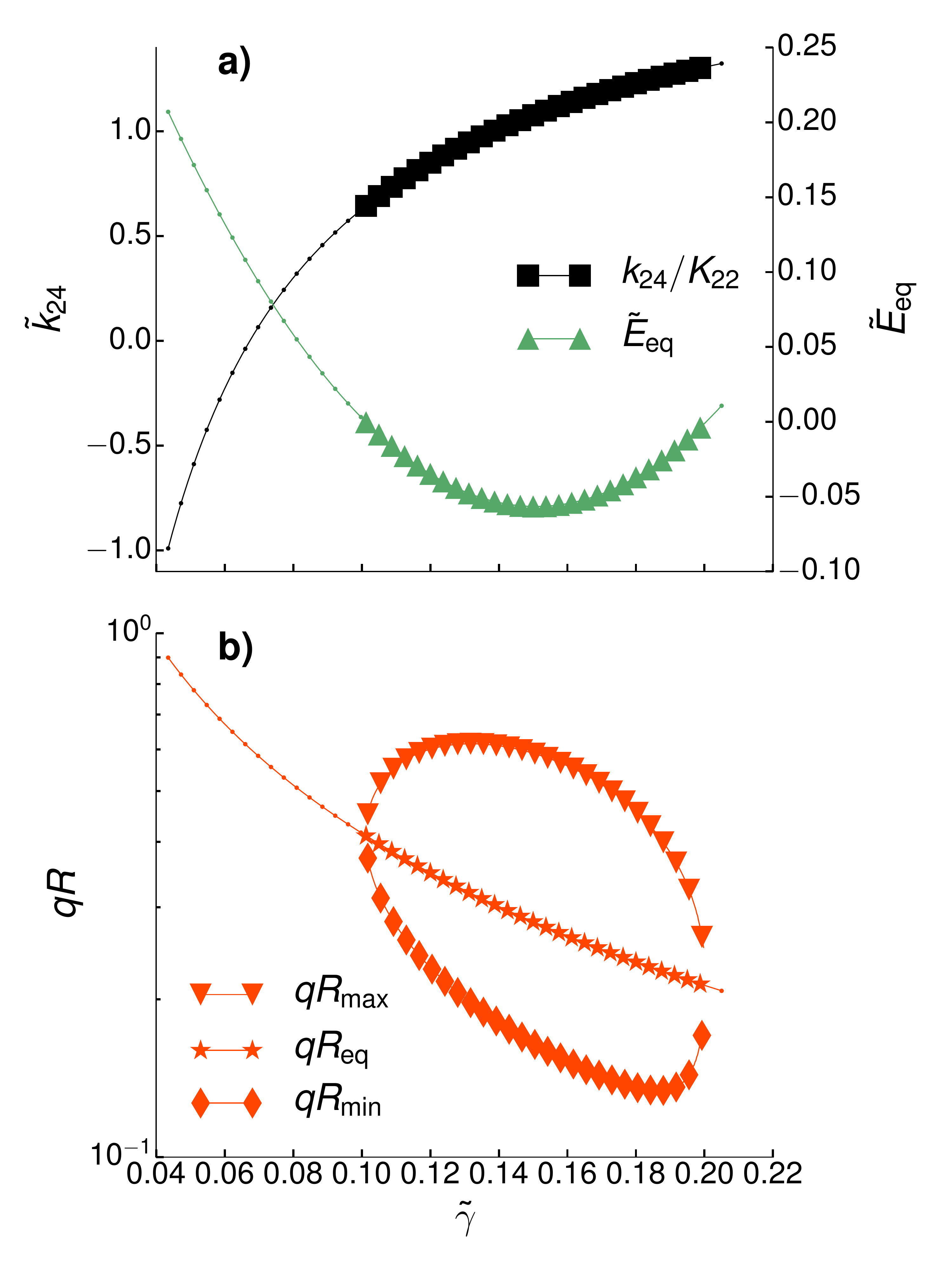}
	\captionsetup{aboveskip = 0pt}
	\caption{Results when the surface twist is restricted to $\psi_{R_{\text{eq}}}=0.31$ (i.e. along the $0.31$ contour in Fig. \ref{fig:PsiRlandscape}) --- the experimentally measured surface twist of corneal fibrils. \protect\subref{fig:E_k24_psiRis0.31}) Reduced saddle-splay $\tilde{k}_{24}$ vs reduced surface-tension $\tilde{\gamma}$ is indicated in black squares and dots, while reduced minimum energy-density $\tilde{E}_{\text{eq}}$ vs $\tilde{\gamma}$ is indicated by green triangles and dots. Dots indicate where fibrils are only meta-stable with respect to the cholesteric phase, and shapes indicate where fibrils are stable with respect to the cholesteric, $\tilde{E}_{\text{eq}}<0$. \protect\subref{fig:qR_psiRis0.31}) The dimensionless fibril radius $q R$ vs $\tilde{\gamma}$. The equilibrium radius that minimizes $\tilde{E}$, $q R_{eq}$, is indicated by stars (when $\tilde{E}_{\text{eq}}<0$) and dots (when $E_{\text{eq}}\geq0$). The minimum and maximum fibril radii that are stable with respect to the cholesteric (i.e. $qR$ values such that $\tilde{E}(q R_{\text{min}})=0$, $\tilde{E}(q R_{\text{max}})=0$ and $qR_{\text{min}}<qR_{\text{eq}}<qR_{\text{max}}$), are indicated by diamonds and triangles, respectively.}
	\label{fig:psiRis0.31}
\end{figure}

\subsubsection{Corneal and other helicoidal fibrils}%%%%%%%%%%%%%%%
For the high surface twist of corneal collagen fibrils, with $\psi_{R_{\text{eq}}} = \SI{0.31}{\radian}$, we show in Fig.~\ref{fig:psiRis0.31} the values of  $\tilde{E}_{\text{eq}}$, $\tilde{k}_{24}$, and $qR$ as a function of $\tilde{\gamma}$. These are determined by calculating the $\psi_{R_{\text{eq}}}=\SI{0.31}{\radian}$ contour line (i.e. $\tilde{k}_{24}$ vs $\tilde{\gamma}$ line) in Fig. ~\ref{fig:PsiRlandscape}, and mapping this relationship onto Figs.~\ref{fig:EnergyLandscape} and \ref{fig:Rqlandscape}, to determine $\tilde{E}_{\text{eq}}$ and $\tilde{qR_{\text{eq}}}$, respectively. Restricting ourselves to thermodynamically stable parameterizations, with $\tilde{E}_{\text{eq}} <0$, from Fig. \ref{fig:psiRis0.31} we expect that $\tilde{\gamma} \in [0.1, 0.2]$, $\tilde{k}_{24} \in [0.6, 1.25]$, and $qR_{\text{eq}} \in [0.2, 0.4]$.

Human corneal fibrils have a typical diameter of $30-35\si{\nano\meter}$ \cite{Schwarz:1953, Hirsch:2001gp, Holmes:2001bt}. We consider a radius of $R\simeq \SI{0.015}{\micro\meter}$ for convenience. This then implies an approximate range of expected chiral wavenumber $q \in [13, 27] \si{\per\micro\meter}$.  This range abuts the expected range from Sec.~\ref{parameters} at larger $qR_{\text{eq}}$, when $\tilde{\gamma} \simeq 0.1$ and $\tilde{k}_{24} \simeq 0.75$ -- this is near point ``3''  of Fig.~\ref{fig:Nonlinearity}.

Using $\tilde{\gamma} \simeq 0.1$ and $q \simeq 13 \si{\per\micro\meter}$, our expected range of $K_{22} \in \left[ {0.6}, {6} \right] \si{\pico\newton}$ from Sec.~\ref{parameters} implies $\gamma \in \left[1.6, 16 \right] \si{\pico\newton\per\micro\meter}$. This is entirely within the expected range of $\gamma \in \left[{0.5}, {25} \right] \si{\pico\newton\per\micro\meter}$. As mentioned, $k_{24}$, is not well constrained --- but nevertheless $\tilde{k}_{24} \simeq 0.75$ is close to the expected scale \cite{Allender:prl1991, Polak:1994iu}.

Corneal fibrils are very close to the stability boundary between fibrils and the cholesteric phase due to their large surface twists. This implies that only a very narrow range of fibril radii are stable with respect to the cholesteric phase, with $\tilde{E}<0$. In Fig.~\ref{fig:psiRis0.31} (b), in addition to $q R_{\text{eq}}$, we indicate the minimum and maximum values for stable fibrils, $qR_{\text{min}}$ and $qR_{\text{max}}$, respectively. For a given $\tilde{\gamma}$ and $\tilde{k}_{24}$, $qR_{\text{min}}$ and $qR_{\text{max}}$ are defined such that $\tilde{E}(q R_{\text{min}})=0$, $\tilde{E}(q R_{\text{max}})=0$ and $qR_{\text{min}}<qR_{\text{eq}}<qR_{\text{max}}$. We see that precisely at $\tilde{\gamma} \simeq 0.1$, there is only a very narrow range of stable fibril radii available for corneal fibrils. Furthermore, a narrow range of corneal fibril radii is observed \cite{Schwarz:1953} and is \textit{required} for corneal transparency \cite{Meek:2015, Maurice:1957}.

In Fig. \ref{fig:PsiRlandscapedifferentK33K22} of Appendix~\ref{K33}, we examine different values of $\tilde{K}_{33}$ to determine whether the correlation between narrow stability and large surface twist is sensitive to our parameter choices. We find that this behaviour persists in a wide range of $\tilde{K}_{33}\in[10,40]$, for $\psi_{R_{\text{eq}}}\simeq\SI{0.31}{\radian}$. From this, we hypothesize that the large surface twist of corneal fibrils may be a result of being at the stability boundary, which in turn is required to narrow the range of accessible fibril radii. Cross-linking after fibrillogenesis could then mechanically stabilize corneal fibrils.

Other ``helicoidal'' or ``C''-type\cite{Ottani:2001} collagen fibrils also exhibit large surface twists with $\psi_{R_{\text{eq}}} \simeq \SI{0.3}{\radian}$ and a narrow unimodal distribution of fibril radii \cite{Raspanti:2018, Ottani:2001, Raspanti:interjbiomacromol1989, Reale:1981}. These helicoidal fibrils are found in e.g. skin, interstitial stroma, and nerve and tendon sheaths. They have a slightly shorter D-period, consistent with the projective coupling hypothesis \cite{Raspanti:2018,Itoh:1996}. Despite their similarity of surface twist, in each tissue helicoidal fibrils exhibit a different unimodal radius -- from   $\SI{0.015}{\micro\meter}$ to $\SI{0.050}{\micro\meter}$ \cite{Raspanti:2018}. Larger radii than seen in corneal fibrils could be accommodated in our model by smaller $q$, or by different points along the stability boundary of Fig.~\ref{fig:PsiRlandscape}. 

Interestingly, some originally helicoidal fibrils from skin that have been disassociated and reconstituted are no longer helicoidal \cite{Brodsky:1980, Gathercole:1987} --- though see \cite{Bouligand:1985}. This implies that fibrillogenesis conditions are important in determining their reduced parameterization; parameters are not simply determined by the molecular type, but also by the environment. While our approach can constrain reduced parameterization with observations of fibril surface twist and radius, a direct assessment of elastic constants within the context of individual fibrils would require different approaches. 

%%%%%%%% Tendon fibrils %%%%%%%%%%%
\begin{figure}[!thb]
	\hspace{-10pt}
	\includegraphics[width=0.5\textwidth]{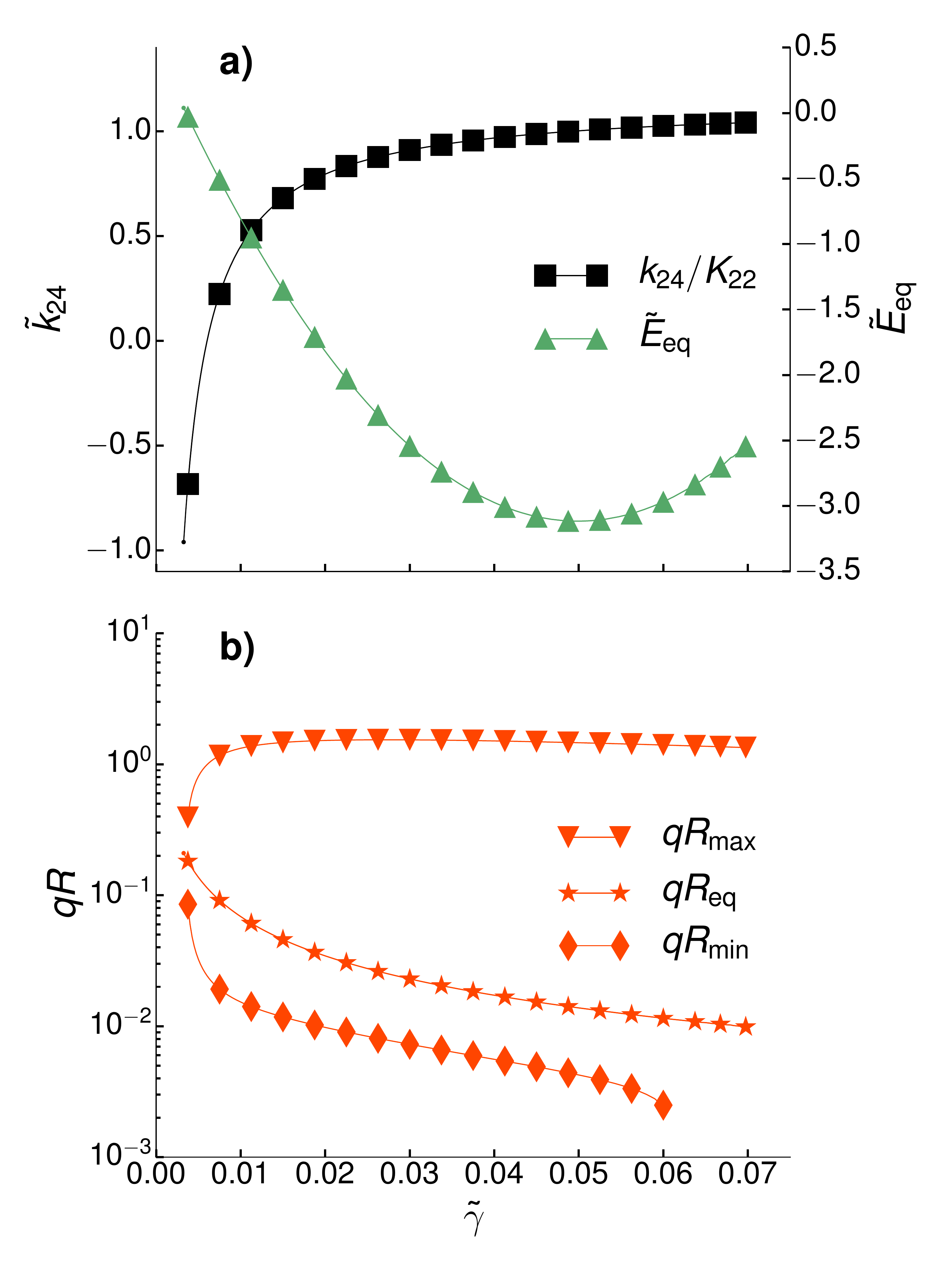}
    {\phantomsubcaption\label{fig:E_k24_psiRis0.1}}
    {\phantomsubcaption\label{fig:qR_psiRis0.1}}
	\captionsetup{aboveskip = -10pt}
\caption{Results when the surface twist is restricted to $\psi_{R_{\text{eq}}}=0.1$ (i.e. along the $0.1$ contour in Fig. \ref{fig:PsiRlandscape}) --- consistent with experimentally observed surface twist of tendon fibrils. \protect\subref{fig:E_k24_psiRis0.31}) Reduced saddle-splay $\tilde{k}_{24}$ vs reduced surface-tension $\tilde{\gamma}$ is indicated in black squares and dots, while reduced minimum energy-density $\tilde{E}_{\text{eq}}$ vs $\tilde{\gamma}$ is indicated by green triangles and dots. Dots indicate where fibrils are only meta-stable with respect to the cholesteric phase, and shapes indicate where fibrils are stable with respect to the cholesteric, $\tilde{E}_{\text{eq}}<0$. \protect\subref{fig:qR_psiRis0.31}) The dimensionless fibril radius $qR$ vs $\tilde{\gamma}$. The equilibrium radius that minimizes $\tilde{E}$, $q R_{eq}$, is indicated by stars (when $\tilde{E}_{\text{eq}}<0$) and dots (when $E_{\text{eq}}\geq0$). The minimum and maximum fibril radii that are stable with respect to the cholesteric (i.e. $qR$ values such that $\tilde{E}(q R_{\text{min}})=0$, $\tilde{E}(q R_{\text{max}})=0$ and $qR_{\text{min}}<qR_{\text{eq}}<qR_{\text{max}}$), are indicated by diamonds and triangles, respectively. Due to the divergent behaviour of the double-twist for $\tilde{k}_{24}\gtrsim1$ (see Fig. ~\ref{fig:EnergyLandscape}), only a small range of $0<\tilde{\gamma}<0.07$ is accessible for $\psi_{R_{\text{eq}}}=\SI{0.1}{\radian}$.}
\label{fig:psiRis0.1}
\end{figure}
% note: what stops us from going to larger tilde{gamma}? Numerical issues?

\subsubsection{Tendon fibrils} %%%%%%%%%%%
For the low surface twist of tendon collagen fibrils, with $\psi_{R_{\text{eq}}} \simeq \SI{0.1}{\radian}$, we show in Fig.~\ref{fig:psiRis0.1} the values of  $\tilde{E}_{\text{eq}}$, $\tilde{k}_{24}$, and $qR$ as a function of $\tilde{\gamma}$. These correspond to mapping the $\psi_{R_{\text{eq}}}=\SI{0.1}{\radian}$ contour line from Fig. \ref{fig:PsiRlandscape} to Figs.~\ref{fig:EnergyLandscape} and \ref{fig:Rqlandscape}, respectively. Restricting ourselves to thermodynamically stable parameterizations, with $\tilde{E}_{\text{eq}} <0$, we expect that $\tilde{\gamma} \in [0, 0.07]$, $\tilde{k}_{24} \in [-1, 1.1]$, and $q R_{\text{eq}} \in [0.01, 0.2]$.  While most of these ranges are larger than those of corneal fibrils, the values of $qR_{\text{eq}}$ for tendon fibrils are significantly smaller. 

Tendon fibrils {\em in vivo}  have a large range of radii, from $\SI{0.02}{\micro\meter}$ to $\SI{0.2}{\micro\meter}$ \cite{Goh:2012, Kalson:2015}, and the distribution varies with age and tissue type. Significantly, fibrils within the same tissue exhibit a broad range of radii.  Nevertheless, the average fibril tendon radius $R = 0.08 \si{\micro\meter}$ from older mouse tails \cite{Kalson:2015} is much larger than typical corneal fibrils. This implies expected values of $q \in [0.13, 2.5] \si{\per\micro\meter}$. These chiral wavenumbers are significantly smaller than for corneal fibrils, but are entirely within the expected range from Sec.~\ref{parameters}.  Combining possible ranges, we then expect the surface tension $\gamma \in [0, 1] \si{\pico\newton\per\micro\meter}$. This is in the lower end of, but largely within, the range expected from Sec.~\ref{parameters}.

However, to have a broad distribution of equilibrium tendon fibril radii within the same section of tissue \cite{Goh:2012,Kalson:2015} would imply a broad range of reduced parameters, and hence of conditions during fibrillogenesis. Tendon fibrils in particular are almost entirely comprised of type-I collagen, and so this variation cannot be attributed to variations of composition. Rather, we believe that non-equilibrium processes are involved in the determination of tendon fibril radii --- as proposed by Kalson et al \cite{Kalson:2015}.

Fibrils with a small surface twist are expected to be quite stable with respect to the cholesteric phase (see Fig.~\ref{fig:EnergyLandscape}). This implies that fibrils at a broad range of different radii around the equilibrium will also be stable with respect to the cholesteric, as shown by the difference in magnitude of $q R_{\text{min}}$ and $qR_{\text{max}}$ in Fig. ~\ref{fig:qR_psiRis0.1}. We note that there is at least a 100-fold range of stable radii available between $R_{\text{min}}$ and $R_{\text{max}}$, with a narrower 5-fold range range between $R_{\text{eq}}$ and $R_{\text{min}}$. The observed 10-fold range of tendon fibril radii fits within the larger range of stable fibrils with respect to the cholesteric.

Our hypothesis then is that non-equilibrium cross-linking works to stabilize fibril radii that are away from $R_{\text{eq}}$, but only have the opportunity to act on fibrils that are stable with respect to the cholesteric (between $R_{\text{min}}$ and $R_{\text{max}}$). Essentially we propose that fibrillogenesis only takes place when fibrils are thermodynamically stable, while cross-linking can freeze (and so prevent) the subsequent slow relaxation of fibril radii towards the minimal energy radius $R_{\text{eq}}$.  We note that this thermodynamic stability may also be of use during remodeling after damage for these load-bearing fibrils \cite{Alves:2016}.

%%%%%%%%%%  CONCLUSIONS %%%%%%%%
\section{Conclusions}
We model collagen fibrils with a double-twist director field of molecular tilt, and identify where a fibril phase is more stable than a cholesteric phase. The stability, dimensionless radius $q R$, and surface twist of the fibrils $\psi(qR)$ are controlled by two dimensionless parameters, the ratio of surface tension to the chiral strength ($\gamma/K_{22}q$) and the ratio of saddle-splay to twist elastic constants ($k_{24}/K_{22}$). The fibril phase is the equilibrium state with respect to the cholesteric phase only when the surface tension is small compared to the chiral strength. Within this limit, the fibril phase can access a wide range of equilibrium configurations ($R_{\text{eq}}$, $\psi(R_{\text{eq}})$). Current experimental observations are consistent with our equilibrium picture, and indicate that controlled equilibrium polymorphism of collagen fibrils may be significant biologically. We suggest that corneal collagen fibrils are formed close to the fibril-cholesteric stability boundary, with large surface twists, in order to achieve a narrow range of fibril radii and to ensure corneal transparency. Conversely, tendon collagen fibrils are formed away from the stability boundary, with small surface twists, but non-equilibrium effects are needed to explain the polydispersity of tendon fibril radii within individual tissues.  A key conclusion is that experimental characterization of a collagen fibril population should always include both radius and surface twist measurements.  

%%%%%%% Appendix %%%%%%%
\appendix
%%%%%%%%%%%%%%%%%%%%%%%%%
\section{Appendix: Dimensional reduction} \label{dimensionless}
The free energy per unit volume of fibril is
\begin{align}\label{eq:freeEnergy}
E =& \frac{1}{R^2}\int_0^Rdr\left[K_{22}r\left(q-\psi'-\frac{\sin2\psi}{2r}\right)^2+K_{33}\frac{\sin^4\psi}{r}\right]\nonumber\\
&-(K_{22}+k_{24})\sin^2\psi(R)+\frac{2\gamma}{R}.
\end{align}
Multiplying eqn. \ref{eq:freeEnergy} by $1/(K_{22}q^2)$ gives the dimensionless free energy per unit volume of fibril,
\begin{align}\label{eq:NoDfreeEnergy}
\tilde{E} =& \frac{1}{\tilde{R}^2}\int_0^{\tilde{R}}d\tilde{r}\left[\tilde{r}\left(1-\tilde{\psi}'-\frac{\sin2\tilde{\psi}}{2\tilde{r}}\right)^2+\tilde{K}_{33}\frac{\sin^4\tilde{\psi}}{\tilde{r}}\right]\nonumber\\
&-(1+\tilde{k}_{24})\sin^2\tilde{\psi}(\tilde{R})+\frac{2\tilde{\gamma}}{\tilde{R}},
\end{align}
where we have defined the dimensionless quantities $\tilde{K}_{33}=K_{33}/K_{22}$, $\tilde{k}_{24}=k_{24}/K_{22}$, $\tilde{\gamma}=\gamma/(K_{22}q)$, $\tilde{r}=qr$, $\tilde{R}=qR$, $\tilde{\psi}(\tilde{r})=\psi(r)$, $\tilde{E}=E/(K_{22}q^2)$.

%%%%%%%%%%%%%%%%%%%%%%%
\section{Appendix: Other $K_{33}$ values} \label{K33}

\begin{figure*}[!tbh]
\centering
\includegraphics[width=0.8\textwidth]{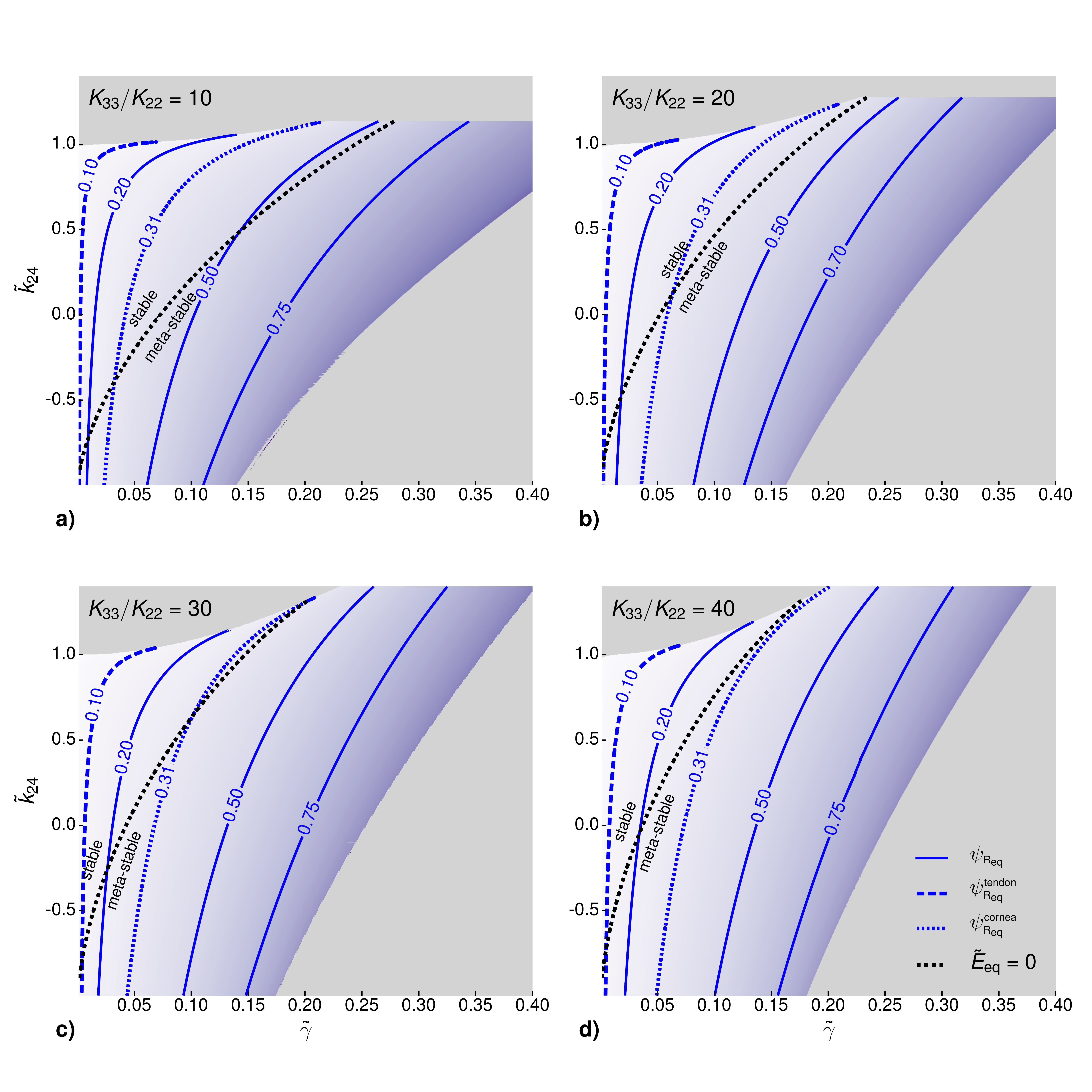}
{\phantomsubcaption\label{fig:K3K2is10}}
{\phantomsubcaption\label{fig:K3K2is20}}
{\phantomsubcaption\label{fig:K3K2is30}}
{\phantomsubcaption\label{fig:K3K2is40}}
\caption{Calculated fibril surface twist $\psi(qR_{\text{eq}})\equiv\psi_{R_{\text{eq}}}$ for different values of $K_{33}/K_{22}$. As $K_{33}/K_{22}$ increases, the surface twist values tend to decrease in size for a given $\tilde{\gamma}$ and $\tilde{k}_{24}$. In \protect\subref{fig:K3K2is10}) the double-twist model predicts the existence of fibrils with very large surface twist, $\simeq\SI{0.52}{\radian}$, which are stable with respect to the cholesteric phase. The surface twist values shown in \protect\subref{fig:K3K2is20} and \protect\subref{fig:K3K2is30} predict a wide range of equilibrium surface twist values (dependent on parameter values), consistent with experimental observations. The $\SI{0.1}{\radian}$ contour, labeled $\psi_{R_{\text{eq}}}^{\text{tendon}}$, is a typical surface twist value of in vivo tendon fibrils. Similarly, the $\SI{0.31}{\radian}$ contour, labeled $\psi_{R_{\text{eq}}}^{\text{cornea}}$, is a typical surface twist value of in vivo corneal fibrils. The surface twist of both fibril types is captured for each $K_{33}/K_{22}$ value shown, but as seen in \protect\subref{fig:K3K2is40}, as $K_{33}/K_{22}\gtrsim40$, the $\SI{0.31}{\radian}$ (cornea) surface twist line transitions completely into the metastable regime. The gray areas in \protect\subref{fig:K3K2is10}-\protect\subref{fig:K3K2is40} correspond to parameter space regions for which no stable or meta-stable double-twist configurations are found.  Note that $\tilde{k}_{24}\equiv k_{24}/K_{22}$, $\tilde{\gamma}\equiv\gamma/(K_{22} q)$, and $\tilde{K}_{33}\equiv K_{33}/K_{22}$.}\label{fig:PsiRlandscapedifferentK33K22}
\end{figure*}

In Fig.~\ref{fig:PsiRlandscapedifferentK33K22} we show the surface twist vs reduced parameters $\tilde{k}_{24}$ and $\tilde{\gamma}$ for a range of $\tilde{K}_{33} \equiv K_{33}/K_{22}$ values: $10$, $20$, $30$, and $40$ for subfigures a)-d) respectively. We note that $\tilde{K}_{33}=30$ corresponds to Fig.~\ref{fig:PsiRlandscape} but is included for ease of reference. 
%%%%%%%%% SURFACE TWIST K33/K22 %%%%%%%%%

%%%%%%%%%%%%%%%%%%%%%%%
\section{Appendix: Power Series Solution} \label{power-series}
We first assume that a convergent power-series expansion in powers of the radius $r$ exists. We then analytically continue this solution to negative $r$, in order to simply note that if $\psi(r)$ is a solution to eqns.~\ref{eqs:BVP} then so is $-\psi(-r)$ --- i.e. $\psi$ is an odd function and will only have odd terms in its power-series expansion.  (This result is independently verified by the numerically relaxed solutions.)   

To simplify our derivation we will use the dimensionless formulation from the previous Appendix~\ref{dimensionless} but will drop the tildes. Then the power series solution for $\psi(r)$ with only odd terms is of the form
\begin{align}\label{eq:genericODDps}
\psi(r)=\sum_{n=0}^{\infty}a_n r^{2n+1},
\end{align}
and satisfies 
\begin{align}\label{eq:appODE1}
(r\psi')'=1+\frac{K_{33}}{2r}\sin2\psi+\frac{(1-K_{33})}{4r}\sin4\psi-\cos2\psi,
\end{align}
where the trigonometric identities $\sin^2x=1/2(1-\cos2x)$ and $\sin2x=2\sin x\cos x$ have been used.  Taylor expanding the trigonometric functions yields
\begin{align}\label{eq:appODE2}
(r\psi')'=&1+\frac{K_{33}}{2r}\sum_{n=0}^{\infty}\frac{(-1)^n2^{2n+1}}{(2n+1)!}(\psi)^{2n+1}\nonumber\\
&+\frac{(1-K_{33})}{4r}\sum_{n=0}^{\infty}\frac{(-1)^n4^{2n+1}}{(2n+1)!}(\psi)^{2n+1}\nonumber\\
&-\left(1+\sum_{l=1}^{\infty}\frac{(-1)^l2^{2l}}{(2l)!}(\psi)^{2l}\right).
\end{align}

The general form of $\psi^n$ in terms of $a_k$ is
\begin{align}
\psi^n&=\left(\sum_{k=0}^{\infty}a_kr^{2k+1}\right)^n\nonumber\\
&=r^n\sum_{k=0}^{\infty}\left(\sum_{j_1+j_2+\cdots+j_n=k}a_{j_1}a_{j_2}\cdots a_{j_n}\right)r^{2k},
\end{align}
where $j_1,j_2,\cdots,j_n \geq 0$ are integer indices and we have used the Cauchy product
\begin{align}\label{eq:cauchy2ps}
\sum_{n=0}^{\infty}a_nx^n\sum_{m=0}^{\infty}b_mx^m=\sum_{k=0}^{\infty}\sum_{l=0}^ka_lb_{k-l}x^k.
\end{align}
Using this we obtain
\begin{align}\label{eq:appODE3}
\sum_{n=0}^{\infty}(2n+1)^2a_nr^{2n}=&\sum_{n=0}^{\infty}c_nr^{2n}\sum_{k=0}^{\infty}p_{2n+1,k}r^{2k}\nonumber\\
&+\sum_{l = 1}^{\infty}d_lr^{2l}\sum_{k=0}^{\infty}p_{2n,k}r^{2k},
\end{align}
where we have defined
\begin{align}\label{eq:c_n}
c_n=\frac{(-1)^n2^{2n}}{(2n+1)!}\left[K_{33}+2^{2n}(1-K_{33})\right],
\end{align}
\begin{align}\label{eq:d_n}
d_l=\frac{(-1)^l2^{2l}}{(2l)!},
\end{align}
\begin{align}\label{eq:p_def}
p_{n,k} = \sum_{j_1+j_2+\cdots+j_n=k}a_{j_1}a_{j_2}\cdots a_{j_n}.
\end{align}
Using eqn. \ref{eq:cauchy2ps}, we re-write eqn. \ref{eq:appODE3}
\begin{align}\label{eq:appODE4}
\sum_{n=0}^{\infty}(2n+1)^2a_nr^{2n}=&\sum_{n=0}^{\infty}\sum_{j=0}^{n}c_{n-j}\,p_{2(n-j)+1,j}r^{2n}\nonumber\\
&+\sum_{n=0}^{\infty}\sum_{k=0}^{n-1}d_{n-k}\,p_{2(n-k),k}r^{2n}.
\end{align}

We can determine each $a_n$ recursively from the eqn. \ref{eq:appODE4}. We find that $a_0=\psi_0'$ is arbitrary. For $n \geq1$, eqn. \ref{eq:appODE4} can be rearranged to give 
\begin{align}\label{eq:a_n}
a_n =  \frac{\sum_{j=0}^{\infty}c_{n-j}\,p_{2(n-j)+1,j}+\sum_{k=0}^{n-1}d_{n-k}\,p_{2(n-k),k}}{[(2n+1)^2-1]},\:\:n\geq1
\end{align}
Since $\,p_{2(n-j)+1,j}$, $\,p_{2(n-k),k}$ depend on all lower coefficients $a_0,\cdots,a_{n-1}$, calculating $\psi(r)$ to high order in $r$ becomes increasingly difficult and is impractical for a broad range of parameters. Nevertheless, we can use the leading cubic  term as a starting point for our numerical relaxation approach: 
\begin{align}\label{eq:psiTOfifthorder}
\psi(r)=\psi_0'r+\frac{(3K_{33}-4)\psi_0'^3-3\psi_0'^2}{12}r^3+\mathcal{O}(r^5)
\end{align}

%%%%%%%%%%% Acknowledgements %%%%%%%%%%%%%
\section{Acknowledgements}
We thank the Natural Sciences and Engineering Research Council of Canada (NSERC) for operating Grants  RGPIN-2013-355291 (LK) and RGPIN-2014-06245 (ADR). SC thanks NSERC and the Nova Scotia Government for fellowship support. 

%%%%%%%%%%%%%%%
\section*{Conflict of interest}
There are no conflicts to declare.

\balance % balance the columns on the final page

%%%REFERENCES%%%
%\renewcommand\refname{Notes and references} % if notes included in references 
\bibliography{Fibril} %You need to replace "rsc" on this line with the name of your .bib file
\bibliographystyle{rsc} %the RSC's .bst file
\end{document}